**Concept Paper**

# Private Blockchain-based Procurement and Asset Management System with QR Code

Alonel A. Hugo
AMA University, Philippines
alonelhugo@gmail.com
https://orcid.org/0009-0004-3987-3500
(corresponding author)

Gerard Nathaniel C. Ngo
AMA University, Philippines

**Abstract**

*Purpose* – The study aims to incorporate private Blockchain technology in the procurement process for the supply office. The procurement process includes the canvassing, purchasing, delivery, and inspection of items, inventory, and disposal. The Blockchain-based system includes a distributed ledger technology, peer-to-peer network, Proof-of-Authority consensus mechanism, and SHA3-512 cryptographic hash function algorithm. This will ensure trust and proper accountability to the custodian of the property while safeguarding sensitive information in the procurement records.

*Method* – The extreme prototyping model will be used as a software development life cycle. It is mostly used for web-based applications and has increased user involvement. The prototype version of the system allows the users to get a better understanding of the system being developed. It also reduces time and cost, has quicker user feedback, missing and difficult functions can be recognized, and confusing processes can be addressed at an early stage.

*Conclusion* – The implementation of a private Blockchain technology has increased privacy, enhanced security, improved efficiency, and reduced complexity over traditional

Blockchain networks. The use of SHA3-512 as a cryptographic hash function algorithm is much faster than its predecessors when cryptography is handled by hardware components. Furthermore, it is not vulnerable to length extension attacks making it reliable in terms of security of data.

*Recommendations* – The study recommends the use of private Blockchain-based technology with the procurement and asset management system in the supply office. The procurement records will be protected against tampering using this technology. This will promote the trust and confidence of the stakeholders.

*Practical Implications* – The implementation of Blockchain technology in developing a system served as advancement and innovation in terms of securing data.

## INTRODUCTION

The role of the Supply Office is to procure all necessary materials needed for the daily operation of an institution. The procurement process is governed by the R.A. 9184 otherwise known as the Government Procurement Reform Act. It is also responsible for the storage and disposal of supplies and equipment of the university. The Supply Office keeps the inventory of property, plant, and equipment (PP&E) and other necessary documents. The whole process of procurement starts from canvassing, procurement, receiving and inspection of delivered items, physical inventory, and disposal of condemned properties.

The conduct of a physical inventory of properties, equipment, and supplies is one way of attesting the physical existence of properties serves as a basis for preparing accounting reports, and is vital in the planning of additional acquisitions. All regular campus personnel are issued with documents called Memorandum Receipts (MR) under their name as custodians of the property and equipment. It can be transferred to another person if the personnel are no longer connected to the office. These records are very important to safeguard until the property or equipment is identified as condemned or unserviceable.

Using Blockchain technology will provide an unchangeable record of all inventory transactions, making it easier to track PP&E and comply with regulations. Blockchain improves security and reduces fraud. Because data stored on the Blockchain is immutable and decentralized, it is highly resistant to tampering and unauthorized access. This helps to prevent counterfeiting and ensures the integrity of the procurement data. It will be easy to find out who did what and when even if any intrusions do happen.

The private Blockchain provides enhanced data security, privacy, and collaboration while facilitating scalability as compared to traditional Blockchain technology. The organization that uses private Blockchain can have faster transaction processing and customizable governance over shared ledger while safeguarding sensitive information.

As technology quickly evolves, there will be a need to have a more highly secure hash algorithm. The SHA3 hash algorithm will be used as a cryptographic hash function. It is much faster than its predecessors when cryptography is handled by hardware components. It is not vulnerable to length extension attacks.

## LITERATURE REVIEW

### *Blockchain Technology*

Several studies have already been conducted and successfully developed a system using Blockchain technology in various applications. For instance, Thejaswini and Ranjitha (2020) developed a system for agricultural production wherein the supply chain transparency and data integration provides certification of genuine food products. The theoretical analysis of the proposed model reveals that the use of Blockchain technology in the food tracking system builds trust among different stakeholders at different stages in the process of agriculture production and provides a hundred percent benefit to them. The study of Islam and Islam (2024) significantly demonstrate the application of Blockchain technology in preventing counterfeit medicine by tracking the records from production until the distribution process using validated blocks of records. Dey et al. (2022) developed a Blockchain-based web system that uses a QR code as a transmission medium that serves as an effective way of transmitting information in different applications such as mobile payment, advertising, passport verification, product traceability, and in other fields.

Because of the technological advancements, Villanueva (2021) exposes the challenges, limitations, and issues of Blockchain technology application. It describes the fundamental elements of Blockchain technology and how policies and regulations limit this technology. It was concluded that identifying these issues at an early stage will allow faster advancement and innovation. Blockchain technology will eventually provide solutions for the current global problems. Kakkar and colleagues (2021) discuss the various implications of Blockchain technology as a competitive advantage by providing a trustworthy and transparent relationship between individuals as well as corporations.

According to Kawaguchi (2019), it has been ten years since Satoshi Nakamoto created Bitcoin and introduced the concept of a Blockchain. The original goal was to propose a solution to the double-spending problem using a peer-to-peer network. Now, Blockchain has proven to have the capacity to deliver a new kind of trust to a wide range of services. Applications as mentioned by Laroiya et al. (2020) describe Blockchain technology as not only redesigning the supply chain management but also the logistics industry, trucking, shipping, freights, and all other modes of transport for transporting goods. Not only all these areas are getting implemented, but also it has reached to government and public sector for various services such as better delivery of services, government-to-public payments, elimination of bureaucracy, prevention of frauds, and many more.

## *Distributed Ledger Technology*

The Department of Trade and Industry Philippines (2019) elaborates on the essential use of Distributed Ledger Technology (DLT) in the transition from a traditional to a digital economy. The DLT enables organizations to have data storage with timestamps and controlled security log access. This technology could transform the delivery of both public and private services with enhanced productivity using various software applications. Bhuvana et al. (2020) explains the use of Blockchain technology in record creation. A distributed record across many computers or digital devices with a process or an activity that cannot be altered retroactively without altering its subsequent processes or activities. It allows a system to own digital goods, assets, and data that can trace the transaction history of everything that is created as a footprint. This way, the Blockchain can be presented as a series of transaction blocks where each block is linked with the previous one through a system of hashes ensuring the interconnection of every new block with the former one. Lakshmi et al. (2021) used the peer-to-peer network as the advantage of having decentralized data storage. The records are made available to everyone in the network minimizing the loss of data when intrusion happens.

## *Using Proof-of-Authority in Private Blockchain*

The use of private Blockchain enables organizations to maintain sensitive information as emphasized by Kakarlapudi and Mahmoud (2021) in a Blockchain-based system. Manolache et al. (2022) explored the use of Proof-of-Authority (PoA) as a consensus mechanism in private Blockchain technology. In comparison to Proof-of-Stake (PoS), PoA has several advantages that focus on the performance and maintenance costs. The PoA consensus algorithm allows easy identification of block creators, increase in accountability, and fixed time intervals. Natoli et al. (2019) suggested that Proof-of-Authority is ideal for small networks like private Blockchains that centralize the overall membership of the Blockchain. The core of PoA is to have a list of validators chosen from trusted nodes.

## *Secure Hash Algorithm*

Lepcha (2023) mentioned the vulnerability of SHA-256 as a common cryptographic hash function algorithm with technological advancements. Lawson (2022) conducted a comparative test between SHA2 and SHA3 algorithms. In the study, it was concluded that SHA2 is slightly faster than SHA3 to a small extent when files are smaller than 40 megabytes. However, SHA3 has a different mathematical composition that increases its potential to avoid collision attacks as compared to SHA2 and previous secure hash algorithms.

## *QR code*

In the study of Agripa and Astillero (2022), QR codes were used in applications across various domains, such as attendance monitoring, recording proof of payment, and authenticating products. Furthermore, QR codes were employed to facilitate access to additional information about a product or item through smartphone scanning. The QR code was used in the study of Shirley et al. (2021) to access the information on passenger luggage in the QR-based Inventory Management System (QR-IMS). It provides easy monitoring, tracking, and fetching of passenger records.

## *Synthesis*

The review of foreign and local related literature of Thejaswini and Ranjitha (2020), Islam and Islam (2024), Dey et al. (2022), Villanueva (2021), Kakkar et al. (2021), Kawaguchi (2019), and Laroiya et al. (2020) showed the significant use of Blockchain technology not only in the global scope but also locally in the Philippines. The application of the technology has evolved from cryptocurrency to a digitalized document routing system. Blockchain technology has now proven to have the capacity to deliver a new kind of trust to the public in the different government services. It proves that Blockchain technology is now being integrated into modern systems.

The related literature of the Department of Trade and Industry Philippines (2019), Bhuvana et al. (2020), and Lakshmi et al. (2021) provided substantial information regarding the use of Distributed Ledger Technology (DLT). The system will implement the use of DLT to provide a tampered-proof record. The users of the system will have identical copies of the procurement documents.

The related literature of Kakarlapudi and Mahmoud (2021) and Natoli et al. (2019) proved that Proof-of-Authority (PoA) is an ideal consensus algorithm for private Blockchain. The use of PoA as a consensus mechanism in the system will help improve the transaction speed and efficiency in energy.

The implementation of SHA3 in the system as discussed in the study of Lepcha, and Lawson (2022) provided a secure one-way cryptographic hash function. The hash value generated by the SHA3 cannot be reconstructed by altering the input without changing the hash. The algorithm will be used to generate a hash value equivalent to the transaction data that aims to ensure the information does not change and avoid fraudulent data.

The studies of Agripa and Astillero (2022), and Shirley et al. (2021) demonstrated that QR code technology can be an effective way of retrieving records in systems. The integration of QR code technology in the system provided better monitoring of supply in an Inventory Management System (IMS).

In conclusion, Blockchain technology provides a reliable, secure, and transparent way for organizations to share data. Its decentralized approach to data storage and sharing, coupled with its immutability, transparency, and security features, make it revolutionary for data sharing in various industries.

## PROPOSED METHODOLOGY

The extreme prototyping model will be used as a model for the software development process. It is one of the types of software prototyping models that is ideal for web-based application development. In software development, AltexSoft (2022) defines software prototyping as an early interactable version of a system that provides user experience to the client. The software prototyping method allows developers of the system to identify user problems earlier to finalize the design before building the actual system. It is composed of six stages, namely: requirements gathering, quick design, prototype development, user evaluation, refinements, and product implementation. Arnowitz et al. (2006) discussed that prototyping allows developers to experiment and easily apply changes without losing much time and effort. This is ideal for developing emerging technology like Blockchain-based systems when user's requirements are not clear.

According to Komatineni (2006), extreme prototyping is used mostly for web-based applications and is a process of three clearly defined stages. The prototype bricks define the different stages of the three levels of prototyping. The first stage involves a static prototype being created, which is just a series of pages. The second stage takes these pages and adds functions using a service layer. The final stage takes this service layer and employs it, implementing the functionality. The prototyping model has increased user involvement in the product even before its implementation as compared to the traditional waterfall model. The prototype version of the system allows the user to get a better understanding of the system being developed. It also reduces the time and cost as the defects can be detected as early as possible. It has quicker user feedback that allows the development team to easily think of a better solution. The missing functionality of the system can be easily identified. The difficult functions can be recognized and confusing processes can be addressed at an early stage.

The ISO/IEC 25010:2023 product quality model will be used as an instrument in evaluating the software quality of the developed system. The ISO (International Organization for Standardization) and IEC (International Electrotechnical Commission) provide worldwide standardization of systems (International Organization for Standardization, 2023). A survey questionnaire based on ISO/IEC 25010:2023 will be given to the participants of the study. It aims to measure the features of the system such as functional suitability, performance efficiency, compatibility, interaction capability, reliability, security, maintainability, flexibility, and safety. This will ensure that the developed system is functioning and follows the standards for software product quality.

## *System Development*

The proposed study aims to develop a web-based system that integrates the use of Blockchain Technology in the procurement process which includes canvassing, procurement, delivery and inspection, inventory, and disposal. It is intended to be used by the employee, canvasser, inspector, property custodian, and administrator of an institution with a Supply Office. The employee can use the system to submit purchase requests for canvassing. The canvasser will be able to process procurement-related documents like the abstract of the canvass and purchase order based on the submitted purchase request. The system will provide a checklist for the delivery of items based on the purchased order for parallel checking. It will allow the inspector to validate the details of the required specifications of the items being delivered. The property custodian will have the ability to generate and print a QR code (Quick Response) for each unit of the property, plant, and equipment (PP&E) after the receiving and inspection phase. The employee will be assigned to be the custodian of the property using the Memorandum of Receipts document (MR) until it becomes unserviceable and for disposal. All transaction data will be recorded using Blockchain and will be distributed in copies across the network for each of the users using Distributed Ledger Technology (DLT). Using Blockchain Technology in the MR documents, the employee can verify the authenticity of the copy and the information specified in the document.

The private Blockchain will be implemented to provide enhanced data security, privacy, and collaboration while facilitating scalability as compared to traditional Blockchain technology. The organization that uses private Blockchain can have faster transaction processing and customizable governance over shared ledger while safeguarding sensitive information.

In a private Blockchain, all user of the system is identified and verified by the system administrator. The computers are connected using peer-to-peer (P2P) network protocol and form a network of nodes. The entire chain of transaction data is distributed and visible to all the nodes in the network using distributed ledger technology (DLT).

The Blockchain is a distributed database of records that are stored in blocks. Each block contains a block hash, timestamped batches of recent valid transactions, a transaction list, and the hash of the previous block. The blocks in the Blockchain follow the parent-child relationship and are linked to one another.

The Poof-of-Authority (PoA) algorithm will be used as a consensus mechanism. It is a reputation-based consensus mechanism that gives the right for a node to generate a new block upon proving the authority to do so. These nodes are called validators, which are limited and chosen by the system administrator. The algorithm of the PoA will assign one primary validator to approve all transactions of one block. Upon reviewing the transaction, the information is assembled into a block.

To provide data security of the blocks, the SHA3-512 (Secure Hash Algorithm version 3) algorithm will be utilized as a Cryptographic Hash Function (CHF). It is the latest addition to the family of SHA or the Secure Hash Algorithm published by the National Institute of Standards and Technology in 2015. The SHA3-512 is a hash function that produces 512-bit digests or hash values of the message. The CHF will generate the hash value to complete the block. The remaining validators will review and confirm the validity of the block created by the primary validator. The new block will be added to the Blockchain upon achieving the consensus. The ledger of the Blockchain record from the database of each node will be updated.

The QR (Quick Response) code functionality will be integrated into the system to facilitate easy identification and tracking of property. It will enhance efficiency in asset management when conducting physical inventory.

Overall, the system will allow the users of the system to digitalize procurement records and validate these records using Blockchain technology in a private setup. The information of these blocks will be protected using the SHA3-512 algorithm. The PoA algorithm as a consensus mechanism will ensure that the authority to validate the transactions is delighted throughout the authorized users of the system.

## CONCLUSIONS AND FUTURE RESEARCH

The use of Blockchain technology in record management systems improves data integrity aside from the security measures that are implemented in a traditional system. A tampered-proof record is an added security detail that any system should utilize. An organization especially in the government sector must adopt this kind of technology to increase credibility and trust towards their stakeholders. The discoveries of the study aim to provide advancement to future researchers that involve the use of Blockchain technology in any type of system that deals with data integrity.

According to the World Economic Forum (2020), there are several legal and regulatory issues that Blockchain Technology is currently facing since many governments and regulators around the world are still working to understand it. These include jurisdiction, technology-neutral regulatory regime, governance and legal documentation, liability, intellectual property (IP), personal data privacy, Decentralized Autonomous Organizations (DAOs), enforceability of smart contracts, and exit from Blockchain. For now, the Blockchain network participants need to ensure that they are compliant with the existing regulations and anticipate the possible changes in the regulatory environment.

The implementation of a private Blockchain in an organization has significant advantages, particularly in addressing the issues of scalability, privacy, control, efficiency, and complexity. The private Blockchain can improve information security's CIA-Triad as discussed by Abasi (2022). In a private Blockchain, the confidentiality requirements are obtainable since access control is enforced which limits the access of information to specific users for specific transactions. It also provides a high level of integrity protection since Blockchain Technology uses a hash function and consensus algorithm. Lastly, it provides a high level of availability since Blockchain Technology uses Distributed Ledger Technology that enables every user to have a full copy of transaction data across the network.

The use of PoA as a consensus mechanism helps improve the transaction speed and it is energy efficient. According to Islam et al. (2022), it is an ideal consensus algorithm for private Blockchain because it requires less computational power since it relies only on a small number of validators. Furthermore, the implementation of SHA3 provides a secure one-way function that makes it a valuable technique in the cryptographic hash function. The hash value generated by the SHA3 cannot be reconstructed by altering the input without changing the hash.

The advantages of using a distributed ledger technology are security, transparency, and decentralization. A distributed ledger is much harder to attack than a centralized database. On the one hand, a decentralized database has no single point of failure (SPOF) since the network can have nodes with their copies of the ledger. On the other hand, the centralized database has the potential risk of SPOF since there is only a single location of the database that could stop the entire operation of the system. Overall, these specifications of Blockchain technology are ideal solutions for organizations, particularly with procurement and asset management systems.

However, some features can be improved for future developments. It is suggested to use a combination of more than one secure hash algorithm to increase the protection against any attacks. It is also important to take note of the network design of the organization to consider different factors that can limit the capability of the system.

In conclusion, the proposed Private Blockchain-based Procurement and Asset Management System offers a solution on how to implement Blockchain technology in a record-keeping system. The future development in this area of research will generate innovation that can be used as solutions to modern world problems.

## PRACTICAL IMPLICATIONS

The implementation of Blockchain technology in developing a system served as advancement and innovation in terms of securing data. As technology quickly evolves, there will be a need to have a more highly secure hash algorithm. The SHA3 hash algorithm will be used as a cryptographic hash function. It is much faster than its predecessors when cryptography is handled by hardware components. It is not vulnerable to length extension attacks. The use of Blockchain technology along with the documents in the procurement and asset management procedures will ensure the integrity of the records. It will provide a genuine record history of the turnover of the property for better accountability. It will promote trust among stakeholders by avoiding fraudulent data.

## ACKNOWLEDGEMENT

The authors would like to express their gratitude towards the people who supported them and contributed to the development of the concept paper. First and foremost, they would like to praise and thank God, the Almighty, who has granted them countless blessings, knowledge, and opportunity to conduct this study. Dr. Jenny Lyn V. Abamo, Dr. Mark Jayson R. Lay, and Dr. Jane Kristine Suarez for the technical expertise they shared since the beginning of the research. Their valuable insights made the success of the study possible. Dr. Cristina M. Signo, campus administrator of the Cavite State University – Carmona Campus, for her undying support towards this academic research. To the faculty and staff of Cavite State University – Carmona Campus: Dr. Regene G. Hernandez, Dr. Brandon G. Sibaluca, Prof. Richard L. Hernandez, Mr. Nestor F. Bueno, Ms. Janine B. Bacosmo, Mr. Jhumel C. Ignas, Mr. Antonino Jose L. Bayson, Mr. St. Joseph M. Lumbog, Mr. Daryl Lyndon T. Supan, Mr. Gerry L. Prado, Mr. Juluis R. Mendoza, Prof. Joe Marlou A. Opella, Prof. Rochelle C. Malabayabas, Ms. Dianna H. Cortez, Mr. Raymond Uminga and others. Their participation in data gathering gives significant development in this study. Finally, to their family and friends for their love and support.

## DECLARATIONS

### *Conflict of Interest*

There is no conflict of interest in the study.

### *Informed Consent*

Not applicable.

### *Ethics Approval*

Not yet applicable.

**Author's Biography**

Alonel A. Hugo is a candidate for the degree in Master of Science in Computer Science with a specialization in Computer Science Theory at the School of Graduate Studies, AMA University, Philippines. He obtained his degree in Bachelor of Science in Computer Science at Cavite State University – Carmona Campus, Philippines in 2011. On the same campus, he has worked as a faculty member under the Department of Industrial and Information Technology handling the BSCS program since 2013. He also serves as the adviser of the Young Programmers and Developers Society (YPADS), a student organization for BSCS students. He has been a member of the Philippine Computer

Society (PCS) and Computing Society of the Philippines (CSP) since 2023, both professional organizations for computing researchers and educators in the Philippines.

Dr. Gerard Nathaniel C. Ngo is an experienced IT professional with in-depth knowledge of technology. He is a graduate of Doctor in Information Technology at the School of Graduate Studies, AMA University, Philippines. In the same school, he obtained his degree in Master of Science in Computer Science in 2015. He took his degree in Bachelor of Science in Computer Science at AMA – Caloocan Campus in 2011. He is currently working as the Head of International Partnerships in the business development unit under ACAD affairs of AMA University, Philippines. Aside from the full-time job, he conducts seminars, trainings, software development projects, network deployment and maintenance, and consultancy in both network and security aspects. He has professional certification in Cisco, AWS, and CompTIA.